# Dynamics of Value-Tracking in Financial Markets

Nicholas CL Beale[a*], Richard M Gunton[b], Kutlwano L Bashe[c], Heather S Battey[d], Robert S MacKay[c]

[a]Sciteb, 23 Berkeley Square, London W1J 6EJ
[b]Department of Accounting, Finance, Mathematics and Economics, University of Winchester, Winchester SO22 4NR
[b]Mathematics Institute, University of Warwick, Coventry CV4 7AL
[c]Department of Mathematics, Imperial College London, London SW7 2AZ
[*]Nicholas.Beale@sciteb.com

# 1. Summary

**The efficiency of a modern economy depends on what we call the Value-Tracking Hypothesis: that market prices of key assets broadly track some underlying value. This can be expected if a sufficient weight of market participants are valuation-based traders, buying and selling an asset when its price is, respectively, below and above their well-informed private valuations. Such tracking will never be perfect, and we propose a natural unit of tracking error, the 'deciblack'. We then use a simple discrete-time model to show how large tracking errors can arise if enough market participants are not valuation-based traders, regardless of how much information the valuation-based traders have. We find a threshold above which value-tracking breaks down without any changes in the underlying value of the asset. Because financial markets are increasingly dominated by non-valuation-based traders, assessing how much valuation-based investing is required for reasonable value tracking is of urgent practical interest.**

# 2. Introduction

The relationship between price and value has been discussed for centuries [1,2], [17]. Financial markets have two main functions in a modern economy: they allow buyers and sellers to transfer ownership of standardised financial assets with well-defined prices and contractual terms, and they provide a "market price" for assets which can then be used for valuing the entire stock of these assets, whether they are traded or not [3]. These valuations in turn influence the behaviour of firms and individuals. A fall in the market value of assets held by a bank or insurance company could result in it becoming insolvent, even without it trading in any of these assets. Changes in the market value of firms have a very strong influence on firm behaviour. In respect of both of these main functions, it is important that market prices should broadly track underlying value [4]. This paper focuses on some market dynamics that could make this difficult.

The essential point is simple. If most of the buy/sell orders in a market come from traders who buy when they think an asset is undervalued and sell when they think it is overvalued, then the equilibrium price of an asset will represent some form of weighted consensus of the valuations by the traders. But if a sufficient volume of orders comes from traders who do not buy/sell on this basis, then market price can substantially diverge from a consensus valuation. And there will generally be a threshold effect, so that when the volume of such non-valuation traders exceeds a certain level, market behaviour changes markedly. Although we demonstrate this point with a very simple model, the underlying point is general. We consider a very stylised market with one asset and three traders. Val buys/sells if the price of the asset is below/above her target price. Mo buys/sells if the price is going up/down. Rand buys/sells at random. We find a threshold ratio of Mo to Val (which depends also on Rand) such that below this ratio the asset price is basically stable and above this it is unstable and liable to crash (Fig. 1(a)). Fig 1(b) shows the probability of crashes in this simple model depending on initial values of Val, Mo and Rand.

The formulation and analysis of such "fundamentalist and chartist" models has a long history [10], regularly showing the destabilising effect of too large a proportion of chartists. Our model exhibits the same phenomenon, but differs in the deterministic part not being differentiable at the equilibrium (so stability analysis is not simply a matter of finding eigenvalues) and in including a stochastic element (Rand).

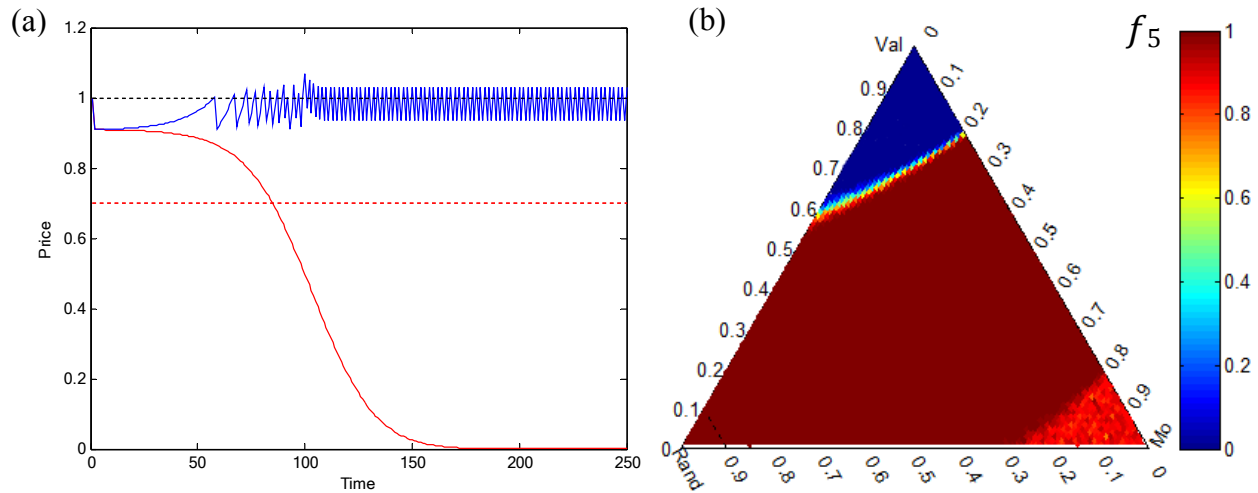

**Fig. 1: Threshold behaviour of a simple simulation model**: (a) price series for a market where the proportion of wealth initially held by a momentum trader (Mo) is 21.5% (blue line) or 21.6% (red line), the remainder being held by a fundamental trader (Val) with constant valuation (black dotted line); (b) probability of a 30% drop in price (5 deciblacks; red dotted line in (a)) over 250 time-steps for different initial proportions of wealth held by Val, Mo or a random investor (Rand). With high enough levels of Mo (right corner), occasionally we get a boom instead of a crash. Ternary plots (here and in Fig. 4) were produced in MATLAB 7.9.0 using package alchemyst [5].

Let's at first assume for simplicity that at any time a given asset has a price $p$ and a value $u$. We propose:

**Definition 1**   $p$ is value tracking within a tolerance $\tau$, or $\tau$-value-tracking, over a certain time period if during all of that period $|\log_2(p) - \log_2(u)| \leq \tau$.

So if $\tau = 0$ the price tracks exactly and if $\tau = 1$ it tracks within a factor of 2. Inspired by Black's proposal about market efficiency [11], we propose to call a deviation of $\tau = 1$ "1 Black" and suggest that a natural unit of tracking is a deciblack which is roughly $\pm 7\%$. A financial crisis is generally associated with a fall in asset prices of 30% or more and if this were due solely to a tracking error this would be roughly 5 deciblacks. Because deciblacks are additive one could also get a similar fall if underlying values fell by 3 deciblacks and the market underpriced assets by a further 2. The time-period in Definition 1 can be arbitrarily short, giving a metric for instantaneous mispricing.

A simple story about asset prices is that a seller S will only sell an asset to a buyer B for a price $p$ which is more than its value $u_S$ to S and less than its value $u_B$ to B and under these conditions $p$ will always be between $u_S$ and $u_B$. If both $u_S$ and $u_B$ are within a ratio $2^\tau$ of the fundamental value $u$ and $p$ fluctuates between them then we may say that $p$ is $\tau$-value-tracking. If S or B have widely differing values for the asset, $p$ may have large fluctuations but so long as $\tau$ is no more than a few deciblacks, the market may appear broadly stable. This may not be the case, however, if there are market participants who are buying and selling on the basis of criteria other than an assessment of value. Major practical examples of users of other criteria in financial asset markets include index funds, which buy and sell assets simply on the basis of their weight in a particular index and whether the fund has net inflows or outflows, some forms of algorithmic traders, and derivatives traders who need to close derivatives positions that have been taken for reasons other than fundamental values of the specific underlying assets. The main thrust of this paper is to demonstrate that such non-valuation-based traders can destabilise a market if their weight becomes excessive compared to the valuation-based traders. We address the question of value-tracking in the framework of a simple model of investing inspired by ecological competition [6].

The view that prices track an asset's value, or a cluster of valuations, is related to the question of informational efficiency of market prices. The claim that all available information is fully reflected in the prices of assets, known as the efficient market hypothesis (EMH) [7], implies that the price moves in line with some function of available information about probable future returns. However, Campbell summarises the situation well: "most economists agree that market efficiency is a useful benchmark but does not hold perfectly" [8]. Additionally, the sense in which prices might "fully reflect" available information has no simple specification. The mapping between an information set and a price can be complicated by heterogeneous and changing beliefs about the interpretation of information sets [9][10], such that one cannot say how a given change in the information would affect the price. The importance of traders who buy an asset when they consider it to be underpriced and sell it when overpriced is not so much that they may happen to abide by some rational economic model of expected future returns as that they are acting with respect to a genuine assessment of its value that is not purely based on previous prices. Without prying into the manner in which they arrive at their valuation, we may take such traders as points of reference for an assessment of value-tracking. We can then formalise the question as to how far prices may reflect the value of an asset, and set up a value-tracking hypothesis – which we do in concluding.

The concept of value-tracking allows prices to be compared to broader assessments of value, such as potential long-term returns under optimal management of an underlying asset, or real value to the economy. Rather than opening up an investigation into various definitions of value, however, for present purposes we define value-tracking in terms of the behaviour of valuation-based traders – i.e. traders who buy an asset when it is priced below what they deem its value to be and sell it when priced above this valuation. We discuss metrics further in Section 4, after looking at a model system.

# 3. Model
## 3.1 Simple Heterogeneous Agent Models

Our work is motivated by the conviction that well-specified simple models can facilitate deeper understanding of the real world. While the extreme complexity of economic reality is beyond human intuition or modelling, we already know that at certain scales and frames of reference, simple patterns may emerge from the interactions of independent actors – such as autocorrelation in prices, or indeed crashes. If we can emulate such distinctive phenomena using simple heterogeneous agent-based models, we may be able to learn something about them without attempting to simulate the complex layers of causality from which they arise [11].

For the question of value tracking, we are interested in traders who act by reference to an exogenous monetary valuation: a judgement of an asset's value based on information about the asset other than its price. Where this valuation comes from and how it is quantified are not critical for present purposes. For the sake of contrast, we then add a class of traders whose beliefs about future returns are purely based on past returns. Two classic types of strategy that imply such contrasting interpretations of price data are fundamental traders and technical traders. A simple model for a fundamental trader specifies a valuation and a rule according to which the trader bids for a certain amount of the asset at each time step when it is undervalued and offers it for sale when overvalued. A simple model for a technical trader, meanwhile, is a momentum trader, who bids for a certain amount of the asset when its price is rising and offers it for sale when the price is falling. These are of course simplistic models, and strongly activist in the sense that they attempt to buy or sell at almost every opportunity, typically in opposition to each other. They are unrealistic in that they allow traders to invest virtually all their cash in a single asset, but they are both rational under certain assumptions, as discussed below. Most importantly, the fundamental version strongly promotes value-tracking, while the momentum version can be parameterised with support from an empirical study of momentum [12]. These two strategies can be taken to represent extremes of a spectrum of more cautious strategies. By combining them in a simple trading model we may simulate some of the features of real markets where diverse participants use many kinds of intermediate strategies.

We next present our simple model, showing how it demonstrates instability thresholds. It operates in discrete time, where we think of the time interval as being one day. Through simulations and analysis we then demonstrate how these depend upon the composition and behaviour of traders in the model. We then refine and extend the model to look for further insights into the behaviour of real economies with multiple valuation-based traders. We end by offering some suggestions for characterising the behaviour of real markets where value-tracking is in doubt.

## 3.2 Val, Mo and Rand

Suppose we have one asset A and a valuation-based trader, Val, who at any time has a stock of the asset $q_V$ and a stock of cash $c_V$ and who trades according to a private valuation $u$. If the price $p$ of the asset is greater than $u$ then Val will try to sell a proportion $k_V^-$ of her holding and if $p$ is less than $u$ then Val will try to buy asset using a proportion $k_V^+$ of her cash. Secondly we introduce Mo, a momentum trader who buys when the asset price is rising and sells when it is falling. Mo uses a short-term weighted average of price changes, called momentum. If momentum is less than 0 then Mo will try to sell a proportion $k_M^-$ of his holding and if it is greater than 0 then he will try to buy asset using a proportion $k_M^+$ of his cash. Thirdly, to allow for unexplained noise and provide better market-making, we introduce Rand, who bids and offers random amounts of his holdings, up to some ceiling. We draw these from uniform distributions on [0, $k_R$] (for selling) and [0,$k_{RC}$] (for buying). Rand can buy and sell at the same time.

These strategies are each rational under certain assumptions. The Val strategy is rational insofar as it expects the asset to tend towards a finite, non-zero price, which would be achieved if this strategy dominated. The Mo strategy is rational insofar as it would be the most profitable strategy if it dominated, even though it would thereby cause the price to crash or boom indefinitely. The Rand strategy could be rational if, hypothetically, the orders were only random with respect to price and in reality reflected the trader's private liquidity or diversification requirements; if Rand truly believes in the EMH then he would reject both information-based and technical motivations for trading (such as used by Val and Mo respectively).

This basic system can be described at any time by the price $p$ of the asset, the valuation $u$, which we hold constant for now, the momentum $m$ of the price, which is updated at each time-step using a rate constant $\mu$ (see below), and the cash and asset holdings of the traders. For Val, Mo and Rand respectively the cash holdings are $c_V$, $c_M$ and $c_R$; and the asset holdings are $q_V$, $q_M$ and $q_R$. We assume for now that the system is closed such that, at all times, $c_V + c_M + c_R = C$ (no inflation) and $q_V + q_M + q_R = Q$ (no new share offerings, buy-backs, etc). We initialise $Q$ as a multiple, $\rho/u$, of $C$, with $\rho$ (the asset–cash ratio) = 4 in the simulations reported here. We also performed analysis for arbitrary $\rho$; the case with $\rho$ <1 produces booms instead of crashes.

We also need a few other fixed parameters and functions:
  i. Commitment parameters as mentioned above, specifying how much of their current holdings a trader will offer when the criteria for buying or selling are met: $k_V^-$, $k_M^-$ and $k_R^-$ for the proportions of their asset holdings that Val, Mo and Rand, respectively, will offer for sale, and $k_V^+$, $k_M^+$ and $k_R^+$ for the proportions of their cash holdings that they will bid at any timestep. We set all these to 10% for the simulations described below. While it would be more realistic to allow the commitment factors to vary with the expected gain, we fix them for simplicity and to understand the dynamics near equilibrium.
  ii. A momentum rate constant determining the weight of the most recent price change in calculating momentum: a factor $\mu$ in the range (0,1), which we set at 0.002 by default. This is used in the momentum update formula $m' = \mu \log(p'/p) + (1-\mu) m$.
  iii. A specification of how the price changes at each interval: we parameterise a power function of the supply/demand ratio (see next section). We also set a maximum price-change $\eta$ of 10% per timestep, similar to daily "limit-up/limit-down" constraints imposed on stocks in real markets in countries including China and the US [13]).

The last two items above are predicated on each time-step representing a day's trading (noting that the constant from Barberis et al. [12] is 0.5 on an annual basis). This reflects our interest in medium to long -term investing behaviour.

## 3.3 Market dynamics

The dynamics of the model depend on the price-updating function. For the simplest model we collect orders at the prevailing price and then, if the sums of purchase and of sale orders are unequal, we update the price using the function: $p' = p (q_p/q_s)^\lambda$, where $q_p$ = total volume of purchase orders and $q_s$ = total volume of sale offers, with $\lambda$ =0.04; this small value ensures that the daily price-change limit of 10% is reached in no more than around 14% of time-steps in runs such as that in Fig. 2c below. In Appendix 1 we show how different price-impact functions change the shape of the dynamics but do not affect the threshold behaviour described below. After the price has been updated, the orders are filled by (i) adjusting sale orders by a factor of the price-change so that the intended amount of cash is still bid, and (ii) where there is an imbalance of amounts bid for versus offered, scaling back the larger of the two to achieve parity. Trades are executed at the updated price.

This approach does not simulate the dynamics of order-filling in real time, where prices tend to move against an order as it is filled, and orders are typically placed (for example as limit orders) in such a way as to minimise the penalties arising from adverse price movements. We also make no attempt to simulate true market clearing, which would require demand curves for each trader; the nature of our Val and Mo strategies makes such parameterisation elusive. We simply note that moving the price by a fractional power of the relative order imbalance, up to a maximum factor which we denote by $\eta$, implies a variable demand function if we assume that the price movement is always sufficient to effect clearing of the market. In general we have an excess supply or demand of $|q_p - q_s|$ and a change in log(price) of $\min(|\lambda.\log(q_p/q_s)|, \log \eta)$, with the opposite sign to the excess demand. If we take this price movement to be sufficient to annihilate the excess of orders, the implied demand function on a scale of log(price) follows the ratio of those two quantities, unless $q_p = q_s$ in which case it is undefined. Clearly this allows a wide range of implicit demand slopes, constrained only to be < 0. Such demand responses are plausible for Val but not for Mo, whose responses to immediate price changes are small but positive. Market clearing considerations thus become most plausible in the presence of random traders, who may be assumed to act as market-makers having a range of negative responses of demand to price.

The benefits of the ratio-power function that we use are that it yields approximately normally-distributed price-changes on a log scale, and that it makes a stability analysis of the model more tractable (see below). An alternative family of price-updating functions is power-law functions based on order imbalances. By making log(price) move by $(q_p – q_s)^\zeta$ for some $\zeta$, up to the limit $\eta$ (Appendix 1), we obtain an implicit demand slope, on a scale of log(price), of $-|q_p – q_s|/\min(\zeta|q_p – q_s|, \log \eta)$ – i.e. $-1/\zeta$ until the price-change limit is reached, at which point it becomes steeper. However, this is still unrealistic for Mo, whose orders should be largely price-insensitive. The effects of using this family of price impact functions are explored in Appendix 1. in earlier simulations we used a somewhat different simplified market mechanism and saw broadly similar behaviour. To be clear: we are not claiming that real financial markets would behave in exactly this way or that any phase transitions would be as sharp as in our model systems, let alone at the same critical values. We are, however, suggesting that analogous phase transitions are likely to occur and that regulators and others should give serious thought to the problems these pose.

## 3.4 Simulation

A basic simulation of our model system, if started in a nonequilibrium state by setting the initial momentum to -0.001, produces a clear threshold effect with increasing initial relative wealth of Mo (Fig. 2).

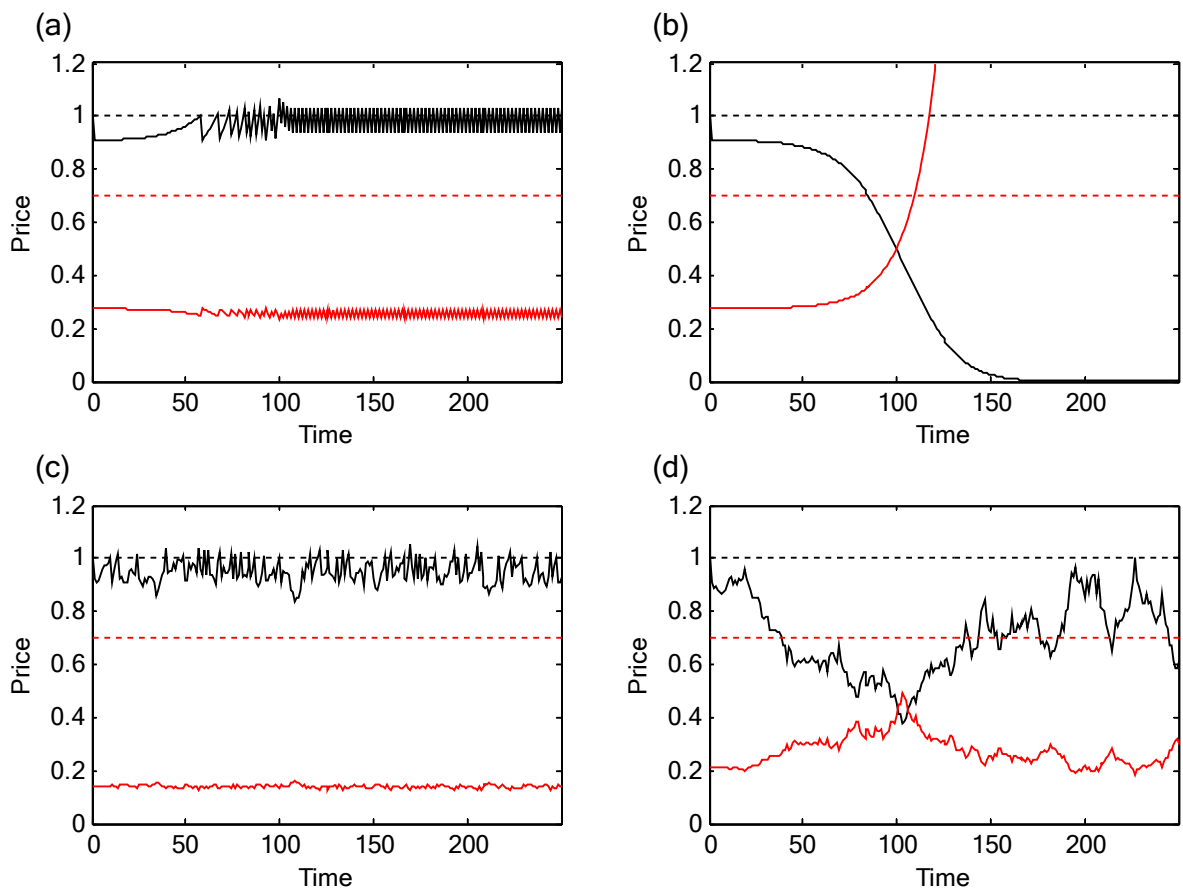

**Fig. 2: Threshold price behaviour from simulations:** with no Rand and initially (a) 21.5% Mo or (b) 21.6% Mo, and with 20% Rand and initially (c) 10% Mo or (d) 14% Mo. Price is indicated by the solid black line, Val's valuation by the dotted black line (1 in each case) and the net worth of Mo relative to Val by the solid red line. Parameters are as specified in Sections 3.2 and 3.3 above, with initial momentum =-0.001. The red dotted line represents a fall of 5 deciblacks.

Fig. 2 demonstrates some important possibilities. First, patterns such as those in Fig. 2(d) would qualify as indicative of a price crash by most criteria (many countries' exchanges have circuit-breakers that halt trading for periods ranging from a few minutes to the remainder of the day if a 10–20% fall occurs [13]). Companies whose stocks moved like this would be considered troubled – and yet, by hypothesis in our model, nothing has changed about the underlying value of the asset. Second, relative outperformance by Mo shifts more resources into momentum strategies. While prices are depressed, Mo outperforms Val, and in the real world we could not expect a valuation trader to maintain a viable position until prices recovered. Indeed, fund managers whose portfolios declined in such a manner could expect to have their funding quickly withdrawn.

## 3.5 Model analysis

The dynamics of this model system in the absence of Rand can be studied analytically. It is somewhat simpler to use the current price to settle the trades rather than the updated price, so we do that. The effect of this change is minimal (Fig. S3). Note that the dynamics are discontinuous where *m*=0 or *p*=*u*, so one cannot analyse stability of the equilibrium *m*=0, *p*=*u*, by linearization.

The state where momentum is zero and the price is equal to Val's valuation (i.e. *m* = 0, *p* = *u*) is an equilibrium, but small deviations could grow and even lead to a crash or a boom ($p \to 0$ or $\infty$). According to our simple ratio-power function, the logarithm of price will change by $\lambda$ times the logarithm of the order imbalance. If *p* departs from *u*, *m* will be non-zero, and there are four regions of possibility:

1. **$p > u, m < 0$**. In this case Val wishes to sell and so does Mo. There are no trades and the price will reduce at the maximum rate allowed until $p < u$: Case 2 below.
2. **$p < u, m < 0$**. Now Val is buying and Mo is selling. The ratio of buy to sell orders determines how the price moves; we write its logarithm as $\alpha = \log \frac{k_V^+ c_V}{k_M^- q_M p}$, where we recall that $k^+$ is the proportion of their cash that a trader will allocate to a buy order and $k^-$ is the proportion of their holding that they will offer to sell, with subscripts V and M indicating Val and Mo respectively. The logarithm of price will change by $\lambda \alpha$ at each step unless capped at $\eta$, with $\alpha$ itself also changing at each step, as a result of both the price-change and the trading (see Appendix 2).
    2.1. If Val's demand is less than Mo's supply we have $\alpha < 0$, so *p* remains < *u* and *m* < 0.
        2.1.1. If $k_V^+ < k_M^-$ then the falling price may eventually cause $\alpha$ to change sign, taking the system to Case 2.2. However, this is not a completely sufficient condition; further details are given in Appendix 2.
        2.1.2. Otherwise, the price will continue to fall, eventually leading to a price crash.
    2.2. If Val's demand exceeds Mo's supply we have $\alpha > 0$, so *p* and hence *m* will increase again. A number of possibilities may follow depending on which of the following happens first:
        2.2.1. If *p* exceeds *u*, the system moves to Case 1.
        2.2.2. If *m* becomes positive, the system moves to Case 3.
        2.2.3. If $\alpha$ becomes negative, the system moves to Case, 2.1.
3. **$p < u, m > 0$**. In this case Val and Mo both wish to buy. Again there are no trades, and the price will increase until $p > u$: Case 4 below.
4. **$p > u, m > 0$**. In this case Val is selling and Mo is buying, and the log-ratio of buy to sell orders can be written as $\beta = \log \frac{k_M^+ c_M}{k_V^- q_V p}$. Similar conditions apply as in case 2.
    4.1. If demand exceeds supply we have $\beta > 0$, so *p* remains > *u* and *m* > 0.
        4.1.1. If $k_M^+ < k_V^-$, then the rising price may eventually cause $\beta$ to change sign, taking the system to Case 4.2 (see Appendix 2 for further detail).
        4.1.2. Otherwise, the price will continue to rise, eventually leading to a price boom.
    4.2. If supply exceeds demand we have $\beta < 0$, so *p* and hence *m* will decrease, and the system may move to Case 3, 1 or 4.1.

For a given set of parameter values, a price crash or boom occurs if Mo is, respectively, selling or buying and the relevant log-ratio of cash to asset holdings has passed zero: $\alpha < 0$ or $\beta > 0$ respectively. A more complete analysis is detailed in Appendix 2, with derivation of sufficient conditions for a price crash or boom in terms of the commitment parameters $k_M^+, k_V^+, k_M^-$ and $k_V^-$, the pricing index λ, and the ratio between the total amounts of cash and asset in the system.

The analysis shows that the conditions for stability concern the overall ratio of asset to cash, the commitment proportions $k_V^+, k_V^-, k_M^+, k_M^-$ of their holdings that each trader will commit at each time-step, and the limit $\eta$ and the parameter λ specifying price movement at each time-step. Fig. 3 demonstrates the effect of the commitment proportions on the initial allocation of wealth to Mo that is sufficient to cause a price crash, comparing results from the analysis with results from simulations. The commitment parameters for Mo and Val are taken to be the same as each other, and the value of the asset at its initial price to be 4 times the quantity of cash in the system. In these simulations, a price drop of 99% within 250 timesteps was taken as a crash (since some parameter combinations produce price trajectories that approach zero but then increase again). We see that, for most commitment levels, Mo need hold only about 25% (orange areas) of the total wealth in the system in order to cause a price-crash. The discrepancies between the two figures reflect the fact that the dynamical analysis provides only sufficient condition for a crash, and turns out to miss the propensity for crashes at low sell commitments. Thus the Mo crash-threshold, as we may call it, only exceeds 40% (green/blue areas) when the buy commitment $k_B$ is less than about 15% or the sell commitment is less than 1%. The relatively low thresholds throughout most of the space are a feature of having the asset–cash ratio greater than 1; in this situation price booms need a relatively high proportion of wealth in Mo's hands (i.e. the Mo boom-threshold is high).

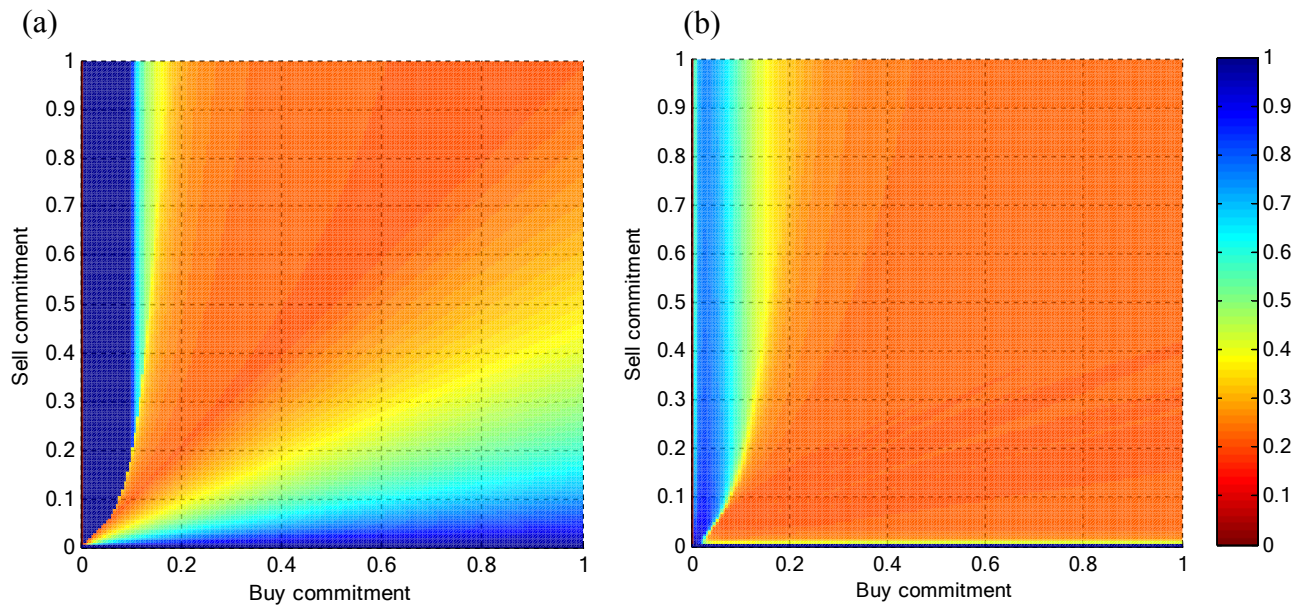

**Fig. 3: Comparison between Mo thresholds: (a) an upper bound obtained analytically and (b) results from simulations**, for a full range of buy (vertical axis) and sell (horizontal axis) commitments. We set the asset–cash ratio to 4, $\lambda = 0.04$, $\eta = 0.1$, momentum constant $\mu = 0.002$, initial price $p_0 = 1$, initial momentum $m_0 = -0.001$. The colour scale indicates how much of the total wealth must be initially held by Mo in order for the price to crash (defined as a fall below 0.01 in the simulations): thus blue means a stable system and orange a very unstable one.

## 3.6 Refining the model

A feature of using simple commitment proportions to determine buy and sell orders in a system with more asset than cash is that there is on average a downward pressure on prices. We see this most clearly when Rand dominates, producing price crashes more often than booms. In the real world this is avoided by such practices as the use of leverage, short-selling and portfolio adjustment. To build a more symmetrical constraint into our simulation model, we make Rand's orders proportional to his overall wealth (as marked to the market price, and subject to having enough cash or asset for a particular order), and specify a proportion of his initial cash and asset holdings as a critical wealth, such that whenever the value of Rand's cash or asset holdings falls below this, his bids and offers are both calculated with reference to the lower of these holdings instead of the overall wealth. Thus when Rand's cash dips below the critical wealth value, his offers of the asset are constrained to the same levels as his bids, and the downward pressure on the price is alleviated. For the simulations below, we set the critical wealth proportion at 20%. We also begin with zero momentum, to avoid seeding conditions for a price crash.

Fig. 4 shows the extent and prevalence of price-drops (relative to the starting value) during 250-time-step simulations for a range of starting conditions. The triangle represents a full range of initial proportions of Val, Mo and Rand, showing that conditions leading to crashes dominate the triangle, and that to avoid them we must start with less than 20% Mo. This threshold is visible as a sharp change in colour either side of the diagonal yellow streak in each plot. In Fig. 4a, the mean relative drop in price jumps from around 0.3 to virtually 1 as this line is crossed. A relative decline of 0.3 is a useful reference point for another way of summarising the results, since most major financial centres have market interventions if an index falls by this amount [13]. We therefore show the probability of a drop of 40% or more in Fig. 4b – and this probability also jumps to 100% as a similar line is crossed. Starting with Mo holding less than 20% of the wealth still has a progressive effect, as average price drops vary from near zero to about 40% and crashes become progressively more frequent   Beyond this, the initial allocation of wealth between Val and Rand in fact matters rather little: 80% Val is sufficient to ensure complete stability, while below the Mo-threshold, 20% Val is generally sufficient to prevent drops in price of more than 30%. The decrease in the severity and frequency of crashes towards runs starting with high levels of Mo (green in right corner of the plots) is due to a tendency for either booms or perfect stability when Mo dominates in the near-absence of Val.

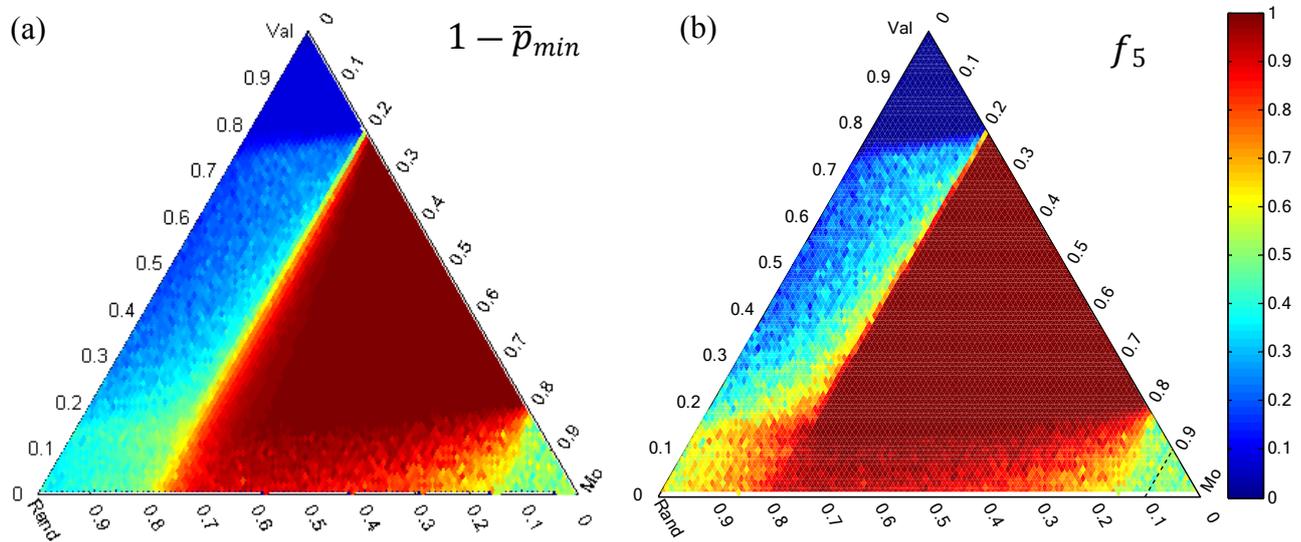

**Fig. 4: Price drops with different initial proportions of Val, Mo and Rand traders: (a) relative price decrease; (b) frequency of a 30% decrease.** The top corner of the triangle represents 100% Val, the right corner 100% Mo and the left corner 100% Rand; a further 5047 starting points are calculated between these extremes, with 100 replicate simulations run from each starting point. Dark blue indicates no decline below 1, while dark red indicates (a) large mean declines or (b) predictable crashes, defined as the frequency of a 30% drop in price (5 deciblacks) over the 100 replicates. All simulations ran for 250 time-steps and used an asset–cash ratio of 4, momentum-smoothing constant of 0.002, buy and sell commitments of 10% and zero initial momentum. Price-drops are defined relative to the starting value, so if the price went up to 1.2 and then down to 0.8 we would record a drop of 20%.

## 4. Multiple valuations and value-tracking

To explore the dynamics of value-tracking and its detection, we must consider multiple valuation-based traders. Our simulation model is readily extended to additional instances of the Val trader, each with their own valuation. We used the refined model (Section 3.6), with parameters as specified above, and specified valuations to be drawn from a gamma distribution with shape and rate parameters equal to 8 (hence a mean, $u$, of 1). We then simulated runs over 1000 time-steps (about four years if each step is a trading day). The results show some interesting features that are suggestive of what might happen in certain real-world situations. In the absence of a momentum trader (Fig. 5) the price tends to move around the range of the valuations – although with considerable fluctuations. With a momentum trader, as before, we tend to witness a price crash (Fig. 6). In both cases, the wealths of the different valuation-based traders progressively diverge. Those that increase the most have valuations closer to the average price and more equality between offers and bids placed; those with extreme valuations tend to move quickly to a position of holding almost entirely cash or asset and see their overall wealth increase the least or in some cases decline.

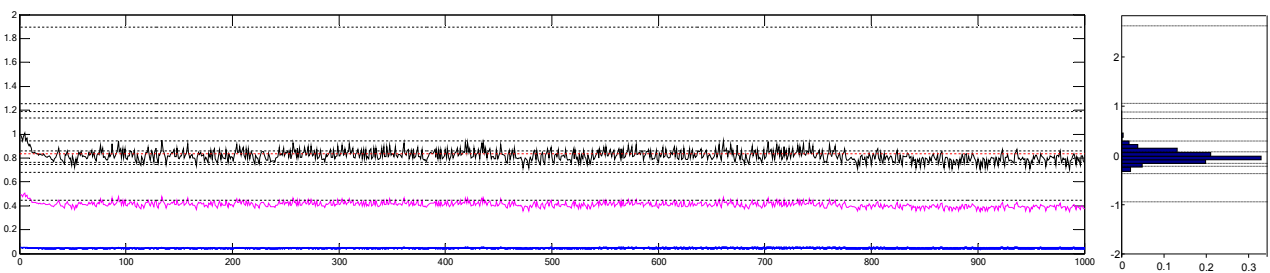

**Fig. 5: Price behaviour from a simulation with 10 valuations and initially 50% Rand and no Mo.** Each of the 10 Val traders initially holds 5% of the wealth (tracked over the course of 1000 time-steps with blue lines), the remainder being with Rand (pink line), whose orders are constrained whenever the cash or asset holding drops below 20% of its initial level. The price is indicated by the solid black line, the valuations by black dashed lines and their sample mean by the red dashed line. The histogram to the right shows, in relative terms, how much time the price spends at different levels with respect to the valuations (grey lines); its vertical axis is calibrated in units of the sample standard deviation among the valuations.

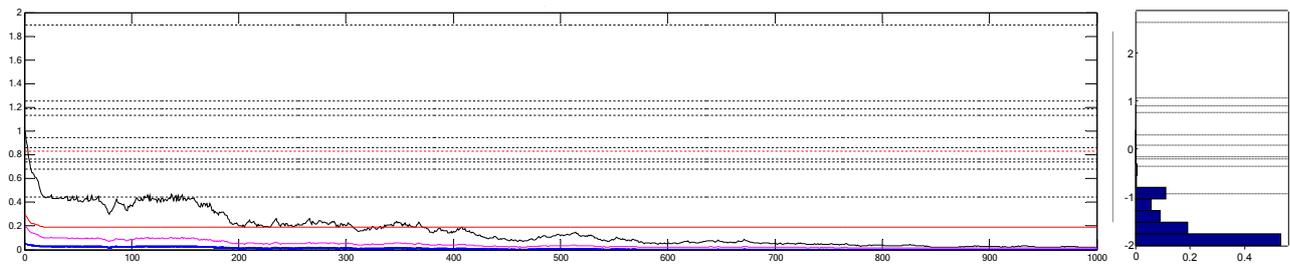

**Fig. 6: Price behaviour from a simulation with 10 valuations, initialised with 20% Rand and 30% Mo.** In the same formats as in Fig. 5, each Val trader initially holds 5% of the wealth, with Mo's wealth additionally indicated by the red solid line. Mo's activity is restricted to the very start of the simulation when he drives the price down, holding almost entirely cash thereafter.

As regards stability, the presence of multiple valuations does not seem to change the interaction among valuation, momentum and random traders. A ternary plot (Fig. S4, Appendix 3) shows similar degrees of price fluctuation when there are 10 valuation traders as found with just one.

The final question is how to quantify $\tau$-value-tracking in such situations. If the valuations are thought of as unbiased estimates $\hat{u}_i$ ($i = 1,...n$) of the true value, then it is sensible to estimate $\tau$ using the appropriately-transformed function of the sample mean of the valuations. See Appendix 4 for details, which also establishes the statistical properties of this estimator. If the valuations are associated with varying degrees of confidence, a weighted average might be used to estimate $u$ and hence $\tau$, with weights inversely proportional to the degrees of confidence, or an appropriate median could be used if some investors' valuations are likely to be biased. This is important where there may be doubt as to whether a particular trader really holds to a particular valuation as stated or inferred by some statistical method, or indeed whether a trader's valuation incorporates value-independent motivations such as a desire to change their position in a holding for strategic reasons. Further options arise if we also have an estimate of the uppermost and lowermost valuations (either because an exhaustive list is available or by some distributional assumption).

# 5. Discussion & Conclusion

Our simple model shows how asset prices may track a fixed valuation, or a consensus valuation, to varying degrees in accordance with the dominance of valuation-based traders over other strategies. The striking threshold effect we observe, whereby sufficient wealth in the momentum trader's account can drive a price crash, also appears in the average of many simulation runs when random investing is included, and with a plurality of valuations. Departures from value-tracking also become progressively greater as the initial wealth of a random trader is increased, but generally without causing a price crash. The exact values of thresholds depend on parameters such as the ratio of asset to cash in the system and the amounts that the traders bid at each time step, but they are insensitive to a change in the price-update function in our investigation. Our result is achieved by combining two strongly proactive trading strategies: both our valuation-based trader and our momentum trader place orders at every opportunity according to their respective investing rule. This contrasts with what may happen if more measured strategies are used, when valuation-based investing does not necessarily produce value-tracking [14].

In our model we observe price crashes and only rarely booms. The analysis shows that this is largely because of having more asset than cash in the system. Our asset–cash ratio of 4:1 is broadly comparable with some recent estimates of the ratio of the combined value of global stock markets to the global monetary base [15] but our model does not account for the important and multi-faceted role of leverage in the economy. It also excludes mechanisms such as liquidations, takeovers and bailouts that may occur when asset prices deviate substantially from fundamentals. But this does not compromise the purpose of this exercise, which is to demonstrate how severe mispricing may occur when valuation-based investment is outweighed by other strategies. With mark-to-market accounting now used for most financial purposes, even transient depressions and bubbles may precipitate irreversible events with repercussions for the real economy and welfare.

Our results are suggestive for the dynamics of current global asset markets, which are characterised by moves away from valuation-based investment. In a market where valuation-based traders predominated, the influence of non-valuation traders might be minimal – and putative technical strategies have often been marginalised as "noise trading" [16]. But the growth of two contrasting types of non-valuation, algorithm-driven investing makes this marginalisation increasingly questionable. High-frequency algorithmic trading typically seeks profits by anticipating movements in price over very short timeframes, typically by identifying transient micro-structures in recent trading on a specific exchange.

Passive asset management, meanwhile, which is also algorithm-driven, is profitable over long timeframes by virtue of simply minimising information costs, in conjunction with sheer market dominance – a dominance that has been growing steadily in recent years around the world, with 48% of Asian, 45% of US and 33% of European equities being held in passive funds at the end of 2017 following recent annual increases of around 3.2, 2.4 and 1.6 percentage points per year respectively [17].  Both high-frequency and passive investment strategies may function as momentum strategies, albeit at different timescales.  High-frequency trading is often based in part on technical analysis of recent price movements; for example it may recognise runs of autocorrelated trades that result from the common practice of breaking a large order into smaller ones.  Passive investing, meanwhile, can create a positive-feedback loop whereby decreases in the price of a stock in an index decrease the weight of that stock in the index, so that outflows of passive funds tend decrease its price further.  Momentum investment can of course drive prices either up or down; our analysis particularly shows the dangers of downward pressure leading to crashes.

The possible existence of thresholds in the market share of non-valuation traders should be a cause for concern for regulators.  With current levels of valuation-based investing, a value-tracking hypothesis may seem plausible, stating that prices of key assets remain within a range of, say, 5 deciblacks of underlying values.  However, there is no guarantee that such stability will persist while the market share of non-valuation-based strategies continues to increase.  Our model suggests that it would be a good idea for financial regulators to estimate the fractions of wealth held by value-traders, momentum-traders and others ('random' traders, including passive managers), and indeed our metric of departures from value-tracking could, with data held by regulators, be applied at the level of individual traders to help build such a picture of market composition.  If prices drop too much or if the fraction of wealth in momentum-traders is getting too large, central banks could have protocols for participation in the market as a value-trader, as has been seen in recent times [HK example].  Value tracking in key assets is important, because when value-tracking fails, market failure may ensue.

**Acknowledgments**
We thank Stephen Acheson, Peter Andrews, Andrew Bailey, Charlie Bean, Alex Brazier, John Y Campbell, Rama Cont, Karen Croxson, Doyne Farmer, Zhao Hao, Jeremy Large, Marcus Miller, Gus O'Donnell, Tom O'Donnell, David Rogers, Alvin Roth and Mike Spence for helpful conversations and suggestions, as well as participants in workshops of the Global Collaboration on Financial System Stability held in London and Hong Kong.  The analysis of the model builds on an MSc group research project report by Kutlwano Bashe, Bhavan Chahal, Nada Jankovicova, Trystan Leng and Chris Norman (2017, University of Warwick).

**Funding Statement**
NB and RG are grateful for support from Aviva, Aberdeen Standard Investments and Capital International. KB is supported by a Botswana Ministry of Education Scholarship. RM is grateful to the Alan Turing Institute for a Fellowship TU/B/000101 that helped enable this collaboration. HB is funded by the U.K. Engineering and Physical Sciences Research Council.

**Data Accessibility**
MATLAB code to run our simulations and generate all the quantitative figures in this paper has been deposited at Dryad and will be accessible at: doi:10.5061/dryad.29tj8r6 [review at https://datadryad.org/review?doi=doi:10.5061/dryad.29tj8r6].

**Competing Interests**
The authors have no competing interests to declare.

**Authors' Contributions**
NB developed the concepts, wrote the original code, directed the work and wrote parts of the paper. RG wrote much of the paper, developed and improved the code, ran most simulations and developed the ternary plots. RM and KB performed the stability analysis and wrote those sections. HB wrote Appendix 4. All authors revised and edited the paper and jointly developed the thinking and exposition.

**Figure captions**

Fig. 1: Threshold behaviour of a simple simulation model: (a) price series for a market where the proportion of wealth initially held by a momentum trader (Mo) is 21.5% (blue line) or 21.6% (red line), the remainder being held by a fundamental trader (Val) with constant valuation (black dotted line); (b) probability of a 30% drop in price (5 deciblacks; red dotted line in (a)) for different initial proportions of wealth held by Val, Mo or a random investor (Rand). With high enough levels of Mo (right corner), occasionally we get a boom instead of a crash.

Fig. 2: Threshold price behaviour from simulations: with no Rand and initially (a) 21.5% Mo or (b) 21.6% Mo, and with 20% Rand and initially (c) 10% Mo or (d) 14% Mo. Price is indicated by the solid black line, Val's valuation by the dotted black line (1 in each case) and the net worth of Mo relative to Val by the solid red line. Parameters are as specified in Sections 3.2 and 3.3 above. The red dotted line represents a fall of 5 deciblacks.

Fig. 3: Comparison between Mo thresholds: (a) an upper bound obtained analytically and (b) results from simulations, for a full range of buy (vertical axis) and sell (horizontal axis) commitments. We set the asset–cash ratio to 4, $\lambda$ = 0.04, $\eta$ = 0.1, momentum constant $\mu$ = 0.002, initial price $p_0$ = 1, initial momentum $m_0$ = -0.001. The colour scale indicates how much of the total wealth must be initially held by Mo in order for the price to crash (defined as a fall below 0.01 in the simulations): thus blue means a stable system and orange a very unstable one.

Fig. 4: Price drops with different initial proportions of Val, Mo and Rand traders: (a) relative price decrease; (b) frequency of a 30% decrease. The top corner of the triangle represents 100% Val, the right corner 100% Mo and the left corner 100% Rand; a further 5047 starting points are calculated between these extremes, with 100 replicate simulations run from each starting point. Dark blue indicates no decline below 1, while dark red indicates (a) large mean declines or (b) predictable crashes, defined as the frequency of a 30% drop in price (5 deciblacks) over the 100 replicates. All simulations ran for 250 time-steps and used an asset–cash ratio of 4, momentum-smoothing constant of 0.002, buy and sell commitments of 10% and zero initial momentum. Price-drops are defined relative to the starting value, so if the price went up to 1.2 and then down to 0.8 we would record a drop of 20%.

Fig. 5: Price behaviour from a simulation with 10 valuations and initially 50% Rand and no Mo. Each of the 10 Val traders initially holds 5% of the wealth (tracked over the course of 1000 time-steps with blue lines), the remainder being with Rand (pink line), whose orders are constrained whenever the cash or asset holding drops below 20% of its initial level. The price is indicated by the solid black line, the valuations by black dashed lines and their sample mean by the red dashed line. The histogram to the right shows, in relative terms, how much time the price spends at different levels with respect to the valuations (grey lines); its vertical axis is calibrated in units of the sample standard deviation among the valuations.

Fig. 6: Price behaviour from a simulation with 10 valuations, initialised with 20% Rand and 30% Mo. As In the same formats as in Fig. 5, each Val trader initially holds 5% of the wealth, with Mo's wealth additionally indicated by the red solid line. Mo's activity is restricted to the very start of the simulation when he drives the price down, holding almost entirely cash thereafter.

**Supplementary Information for Revised Manuscript**

# Dynamics of Value-Tracking in Financial Markets

*Nicholas CL Beale, Richard M Gunton, Kutlwano L Bashe, Heather S Battey, Robert S MacKay*

**Contents:**
Appendix 1: Comparison of different price-impact functions
Appendix 2: Dynamical analysis of simple model
Appendix 3: Summary of simulations with multiple valuations
Appendix 4: Statistical properties of an estimator for τ

**Appendix 1: Comparison of different price-impact functions**

Comparing the power-ratio price update function used for the main analyses with two instances of an alternative power-law function yields similar dynamics, with the same threshold for the initial amount of Mo required to drive a price crash. The power-law function for updating price *p* to price *p'* is based on the difference between buy and sell orders at a given time rather than their ratio; it is defined as:

$$p' = p \, |(A-O)/\lambda|^{\zeta} * \text{sign}(A-O)$$

where:

| | | |
|---|---|---|
| A | ≜ | total volume of purchase orders at the time |
| O | ≜ | total volume of sale offers at the time, |
| λ | ≜ | a liquidity parameter (here kept at 1) |
| ζ | ≜ | a parameter representing the nonlinearity of the impact function. |

We look at a so-called log-linear version [13] in which $\zeta = 1$ and a non-linear version where $\zeta = 0.8$. In the graph headings below:

| | | |
|---|---|---|
| Mo | = | the initial % of all Assets and Cash held by Mo. |
| Rand | = | the initial % of all Assets and Cash held by Rand |
| Z | = | ζ (see above) |
| $m_1$ | = | initial momentum (always -0.001) |

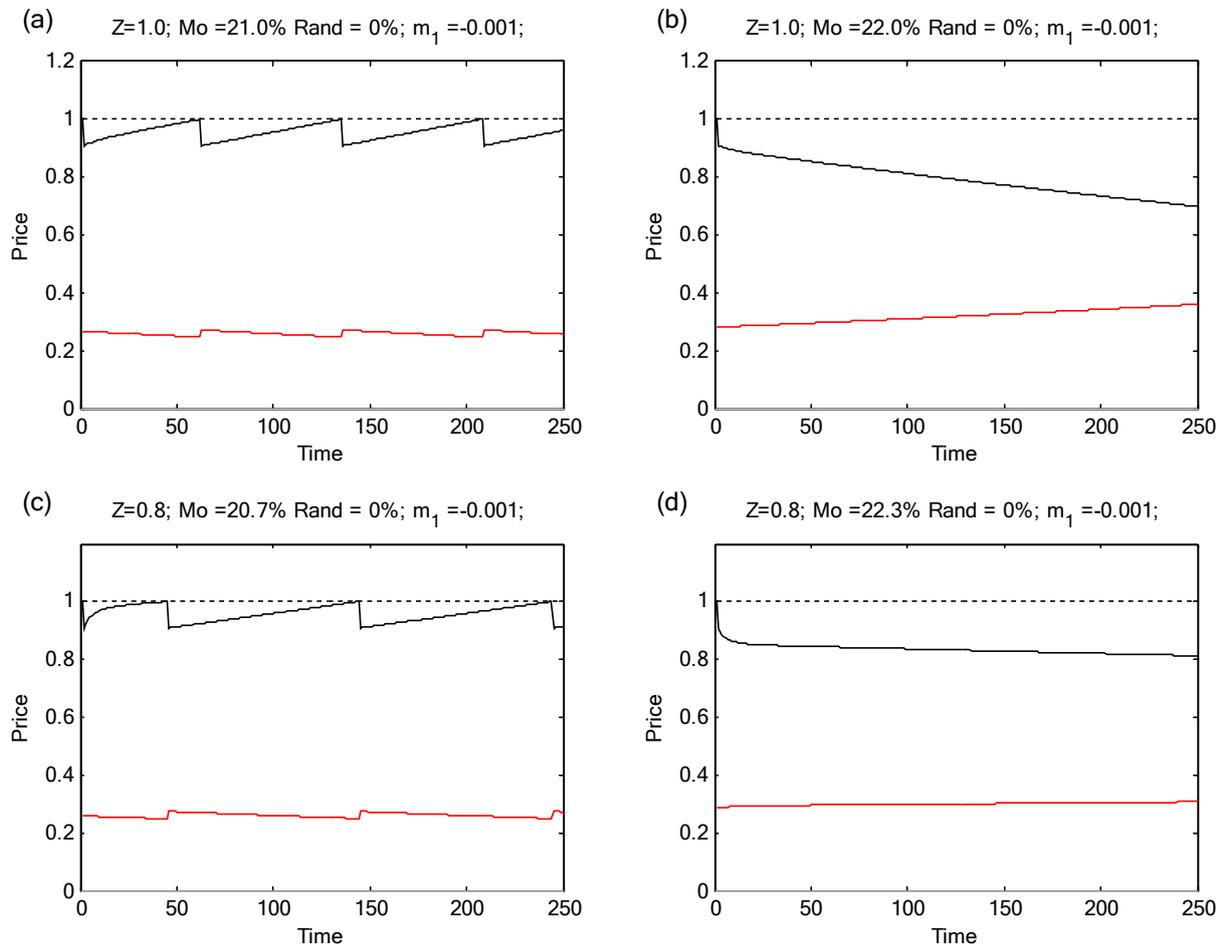

**Fig. S1: Threshold dynamics of price (black line) and relative wealth allocation (red line showing Mo:Val ratio)**, showing threshold effect (left versus right column) around 21.5% Mo: first with a log-linear function ($\zeta = 1$) starting at (a) 21% Mo and (b) 22% Mo; and then with a power-law price-impact function ($\zeta = 0.8$) starting at (c) 20.7% Mo and (d) 22.3% Mo. The threshold effects are less pronounced than in the case shown in the main text (see Fig. 2), but the levels are unchanged. Other parameter values are as laid out in section 3.2 above.

A market-clearing rule has been found to produce similar results (Scholl and Farmer, private communication).

## Appendix 2: Dynamical analysis of simple model

This analysis pertains to the model with Val and Mo trading at current prices, and no Rand. At any given time, Val holds quantity $q_V$ of asset and amount $c_V$ of cash. The current price is $p$ and Val's target price is $u$. If $p > u$ then Val wishes to sell quantity $k_V^- q_V$, or if $p < u$ then Val wishes to spend $k_V^+ c_V$ to buy quantity $k_V^+ c_V / p$. If $m > 0$ then Mo wishes to spend $k_M^+ c_m$ to buy quantity $k_M^+ c_M / p$. If $m < 0$, Mo wishes to sell quantity $k_M^- q_M$.

Purchase and sale orders are matched as far as possible, trading at the current price $p$. The price is then updated as:

1. $p' = p(q_P/q_S)^\lambda$

where $q_P$, $q_S$ are the quantities of purchase and sale orders respectively and $\lambda$ is a 'liquidity' parameter set at 0.04, but $p'$ is capped to the edges of $[pe^{-\eta}, pe^{\eta}]$ if the above formula makes it go outside.

The momentum is updated by:

2. $m' = \mu \log \frac{p'}{p} + (1-\mu)m$

Note that the total amounts of cash and of asset are conserved:

3. $c_V + c_M = C$
4. $q_V + q_M = Q$

So we can reduce the four state variables to:

5. $\tilde{c}_V \triangleq \frac{c_V}{C}$
6. $\tilde{q}_V \triangleq \frac{q_V}{Q}$

Let:

7. $A \triangleq \frac{k_M^- uQ}{k_V^+ C}$
8. $B \triangleq \frac{k_V^- uQ}{k_M^+ C}$
9. $\pi \triangleq \log \frac{p}{u}$
10. $\alpha \triangleq \log \frac{k_V^+ c_V}{k_M^- q_M p} = \log \frac{\tilde{c}_V}{1-\tilde{q}_V} - \log A - \pi$
11. $\beta \triangleq \log \frac{k_M^+ c_M}{k_V^- q_V p} = \log \frac{1-\tilde{c}_V}{\tilde{q}_V} - \log B - \pi$

We can reconstruct $c_V, c_M, q_V, q_M$ from $C, Q, \alpha, \beta, \pi$ (apart from on the back-diagonal $\tilde{c}_V + \tilde{q}_V = 1$, where $\alpha$ and $\beta$ are constant) by:

12. $\tilde{q}_V = \frac{e^{-\pi} - Ae^\alpha}{Be^\beta - Ae^\alpha}$
13. $\tilde{c}_V = Ae^\alpha \frac{Be^{\beta+\pi} - 1}{Be^\beta - Ae^\alpha}$

Note that $\tilde{q}_V, \tilde{c}_V \in [0,1]$ imposes some constraints on $\alpha$ and $\beta$. Specifically, each is satisfied iff

14. $(\beta + \pi + \log B)(\alpha + \pi + \log A) \leq 0$.

We can now work out the dynamics in terms of the variables $m, \pi, \alpha, \beta$. $\alpha$ measures Val's buying power relative to Mo's selling power; $\beta$ measures Mo's buying power relative to Val's selling power.

**Case 1:** $\pi > 0, m < 0$:

15. $\pi' = \pi - \eta < \pi$
16. $m' = (1-\mu)m - \mu\eta < 0$
17. $\alpha' = \alpha + \eta$
18. $\beta' = \beta + \eta$

So $m$ remains negative and eventually $\pi < 0$: go to Case 2.

**Case 2:** $\pi < 0, m < 0$:

Let

19. $\phi = \begin{cases} \lambda\alpha & \text{for } |\alpha| \leq \eta/\lambda \\ \text{sign}(\alpha)\eta & \text{for } |\alpha| \geq \eta/\lambda \end{cases}$

**2a.** If $\alpha < 0$:

20. $\pi' = \pi + \phi < \pi$

21. $m' = (1-\mu)m + \mu\phi < 0$

22. $\alpha' = \alpha - \phi + \log \dfrac{1 - k_V^+}{1 - k_M^- e^\alpha}$

23. $\beta' = \beta - \phi + \log \dfrac{(e^{-\alpha} - Ae^\pi) + k_V^+ A(e^\pi - B^{-1}e^{-\beta})}{(e^{-\alpha} - Ae^\pi) + k_M^-(Be^{\beta + \pi} - 1)}$

**2b.** If $\alpha > 0$:

24. $\pi' = \pi + \phi > \pi$

25. $m' = (1-\mu)m + \mu\phi > m$

26. $\alpha' = \alpha - \phi + \log \dfrac{1 - k_V^+ e^{-\alpha}}{1 - k_M^-}$

27. $\beta' = \beta - \phi + \log \dfrac{(1 - Ae^{\alpha + \pi}) + k_V^+ A(e^\pi - B^{-1}e^{-\beta})}{(1 - Ae^{\alpha + \pi}) + k_M^-(Be^{\beta + \pi} - 1)}$

Note that the formulae for $\pi'$ and $m'$ are the same in the two cases, but the resulting inequalities are different. Note also the nice feature that the update to $\alpha$ does not depend on any of the other variables. The graph of $\alpha'$ as a function of $\alpha$ is continuous and has slope at least $1 - \lambda > 0$ (except at $0, \pm \eta/\lambda$, where the slope is undefined), so the dynamics is monotone (if it starts by increasing it continues to increase; if it starts by decreasing then it continues to decrease), but depending on the parameters the graph can have various forms, in particular with different numbers of fixed points (i.e. values at which $\alpha' = \alpha$).

Suppose $k_V^+ \geq k_M^-$. Then $\alpha'(0) \leq 0$. But if $\alpha$ starts negative then it remains negative, the conditions $\pi < 0$, $m < 0$ are preserved and $\pi$ goes to $-\infty$ (crash!). If $\alpha$ starts positive then $\pi$ or $m$ may turn positive and we transition to Case 1 or Case 3 respectively. In particular, if there is a positive fixed point $\alpha_+$ and $\alpha$ starts larger than $\alpha_+$ then one of $\pi$ and $m$ is forced to become positive eventually. If $\alpha$ starts smaller than all the fixed points (or there are none) then $\alpha$ may become negative before either of these happen and again we get a crash. To decide the outcome, compare:

28. $\pi_n = \pi_0 + \sum_{k=0}^{n-1} \phi_k$

29. $m_n = (1-\mu)^n m_0 + \mu\sum_{k=0}^{n-1}(1-\mu)^{n-k-1}\phi_k$

for increasing $n$ until one becomes positive (if this never happens, we get a crash). We exit Case 2 with $\pi_n \leq \eta$, $m_n < \mu\eta$.

Suppose instead that $k_V^+ < k_M^-$. Then $\alpha'(0) > 0$. If there is a negative fixed point $\alpha_-$ and $\alpha$ starts less than or equal to this then $\alpha$ remains negative and we get a crash. Otherwise $\alpha$ eventually turns positive and remains so, and then one of $\pi$ and $m$ eventually becomes positive and we transition to Case 1 or 3 respectively. The outcome is decided as above.

During the time in Case 2, $\alpha$ and $\beta$ evolve. If there is no crash, then where they end up is the main determinant of what happens next. For example, if $k_V^+ > k_M^-$ and $\alpha$ starts less than the positive fixed points and the next transition is to Case 1, then $\alpha$ decreases (except for the increases in Case 1) and there is more chance that next time round in Case 2 it will go negative and elicit a crash. Thus the conditions outlined above are sufficient for a crash but may not be necessary.

**Case 3:** $\pi < 0, m > 0$:

This is the same as Case 1 with the sign of $\eta$ changed and $m$ remaining $> 0$. It goes eventually to Case 4.

**Case 4:** $\pi > 0, m > 0$:

Let

30. $\psi = \begin{cases} \lambda\beta & \text{for } |\beta| \leq \eta/\lambda \\ \text{sign}(\beta)\eta & \text{for } |\beta| \geq \eta/\lambda \end{cases}$

**4a.** If $\beta > 0$:

31. $\pi' = \pi + \psi > \pi$

32. $m' = (1 - \mu)m + \mu\psi > 0$

33. $\beta' = \beta - \psi + \log \frac{1 - k_M^+ e^{-\beta}}{1 - k_V^-}$

34. $\alpha' = \alpha - \psi + \log \frac{(1 - Be^{\beta + \pi}) + k_M^+ B(e^\pi - A^{-1}e^{-\alpha})}{(1 - Be^{\beta + \pi}) + k_V^- (Ae^{\alpha + \pi} - 1)}$

**4b.** If $\beta < 0$:

35. $\pi' = \pi + \psi < \pi$

36. $m' = (1 - \mu)m + \mu\psi < m$

37. $\beta' = \beta - \psi + \log \frac{1 - k_M^+}{1 - k_V^- e^\beta}$

38. $\alpha' = \alpha - \psi + \log \frac{(e^{-\beta} - Be^\pi) + k_M^+ B(e^\pi - A^{-1}e^{-\alpha})}{(e^{-\beta} - Be^\pi) + k_V^- (Ae^{\alpha + \pi} - 1)}$

Again, $\beta'$ depends only on $\beta$ and is continuous and of slope at least $1 - \lambda > 0$.

Suppose $k_M^+ \leq k_V^-$. Then $\beta'(0) \geq 0$. If $\beta$ starts positive then it remains so, $\pi$ and $m$ remain positive and $\pi$ goes to $+\infty$ (boom!). If $\beta$ starts negative then $\pi$ or $m$ may turn negative and we transition to Case 3 or 1, respectively. In particular, if there is a negative fixed point $\beta_-$ and $\beta$ starts less than this then one of $\pi$ and $m$ is forced to become negative eventually. If $\beta$ starts larger than all the negative fixed points (or there are none) then $\beta$ may become positive before either of these happens and then we again get a boom.

Suppose instead that $k_M^+ > k_V^-$. Then $\beta'(0) < 0$. If there is a positive fixed point $\beta_+$ and $\beta$ starts larger than $\beta_+$ then $\beta$ remains positive and we get a boom. Otherwise $\beta$ eventually turns negative and then one of $\pi$ and $m$ eventually becomes negative and we transition to Case 3 or 1 respectively. As before, the conditions outlined here are sufficient for a boom but may not be necessary.

To analyse the stability of $\pi = 0$; $m = 0$ completely, we need to keep track of how much $\alpha$, $\beta$ change in cycles through the different cases and whether we get pushed into crash or boom scenarios. $\pi = 0$, $m = 0$ is a subset of the plane parametrised by $\alpha$, $\beta$, constrained by (14) (with $\pi$ near 0 for the stability analysis) and the stability will depend on $\alpha$; $\beta$, as well as the parameters like $k$, $\mu$, $\lambda$, $\eta$ and $A$, $B$.

This is a non-standard stability problem, because the map is discontinuous at $\pi = 0$; $m = 0$. But we will understand stability as meaning that $\pi$ and $m$ do not make large excursions from 0 on the scale of $\eta$, $\mu\eta$. Each of the following are then sufficient conditions for a crash for $\pi$, $m$ starting respectively within $\eta$, $\mu\eta$ of 0:

- $k_V^+ \geq k_M^-$ and $\alpha < -\eta$
- $k_V^+ < k_M^-$ and $\alpha < \alpha_- - \eta$ (where $\alpha_-$ is the largest negative fixed point in Eq.22)

and sufficient conditions for a boom:

- $k_M^+ \leq k_V^-$ and $\beta > \eta$
- $k_M^+ > k_V^-$ and $\beta > \beta_+ + \eta$ (where $\beta_+$ is the smallest positive fixed point in Eq.33)

but there are likely to be other cases going to crash or boom too.

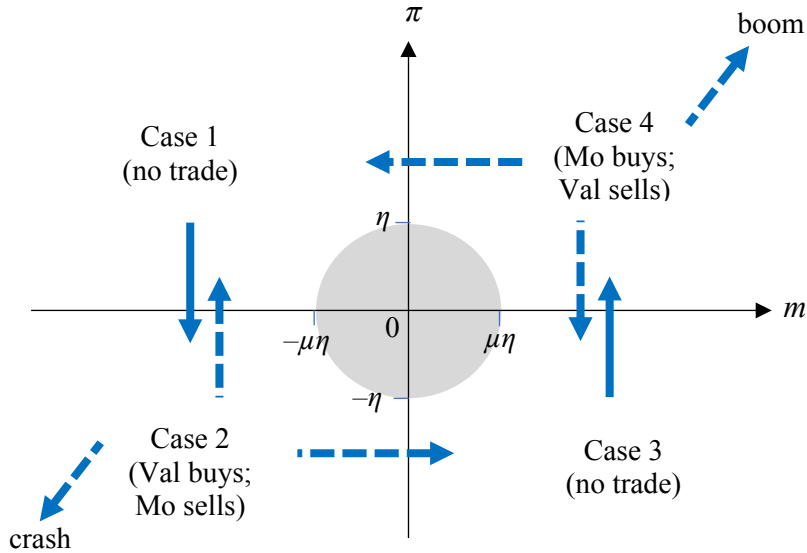

**Fig. S2: Phase space illustration for the stability analysis**, with arrows indicating the possible increases or decreases on each axis within each quadrant (not to scale) according to values of discriminating parameters. The grey ellipse indicates the zone within $\eta$, $\mu\eta$ of 0 where we consider trajectories to start, and from which we might consider any large departures to count as indicating instability.

To derive sufficient conditions for a crash, we need to find the fixed point $\alpha_-$. For $k_V^+ \geq k_M^-$ we can simply use $\alpha_- = 0$. For the case $k_V^+ < k_M^-$, we refer to Eq. 22, which has at most 3 fixed points. Finding these means solving for $\alpha' = \alpha$ here, which is equivalent to finding the roots of

39. $f(\alpha) = -\phi + \log \dfrac{1 - k_V^+}{1 - k_M^- e^\alpha}$

For the range $\alpha \in (-\infty, -\eta/\lambda)$, $\phi = -\eta$ (from Eq. 19), so this can be solved exactly to get the fixed point

40. $\alpha_- = \log \dfrac{1 - e^\eta(1 - k_V^+)}{k_M^-}$

which exists iff

41. $1 - e^{-\eta} < k_V^+ < 1 - e^{-\eta} + k_M^- e^{-\eta\left(1 + \frac{1}{\lambda}\right)}$

For the range $\alpha \in [-\frac{\eta}{\lambda}, 0]$, $\phi = \lambda\alpha$ (again from Eq. 19). Here Eq. 39 cannot be solved exactly and numerical methods are needed. Because we have $k_V^+ < k_M^-$, $f(0) > 0$ and $f'(0) > 0$ if $k_M^- > \frac{\lambda}{1+\lambda}$ so the roots exist if the minimum of $f$ over $[-\frac{\eta}{\lambda}, 0]$ is negative. The minimum is at

42. $\alpha_{min} = \log \dfrac{\lambda}{k_M^-(1+\lambda)}$

so provided $f(\alpha_{min}) < 0$, there will be 2 roots. Using the Newton–Raphson method starting with $\alpha_0 = -0.01$, we can find the larger of these two roots, which is the desired one. If no roots exist in the range $\alpha \in [-\frac{\eta}{\lambda}, 0]$ then the one in $(-\infty, -\eta/\lambda)$, if it exists, is the required root.

Once we have the desired fixed point $\alpha_-$, if it exists, we note that

43. $\alpha \leq \alpha_- - \eta \Rightarrow \log \dfrac{k_V^+ c_V}{k_M^- q_M p} \leq \alpha_- - \eta$

Finally, for an exploration of a subset of the parameter space, let $\rho \triangleq Qu/C$ and define the proportion of wealth held by Mo as:

44. $\theta = \dfrac{c_M + q_M \rho}{C + Qp}$.

Then we have $c_V = (1 - \theta)C$, $q_M = \theta \rho C$. If we set $u = 1$ and have $p = 1$ initially, Eq. 43 can be rearranged to give an inequality on $\theta$ that is sufficient to lead to a crash:

45. $\theta \geq \dfrac{k_V^+}{k_M^- \rho e^{\alpha_- - \eta} + k_V^+}$

In the cases where a fixed point for $\alpha$ does not exist, we set $\alpha_- = -\infty$ to give $\theta \geq 1$.

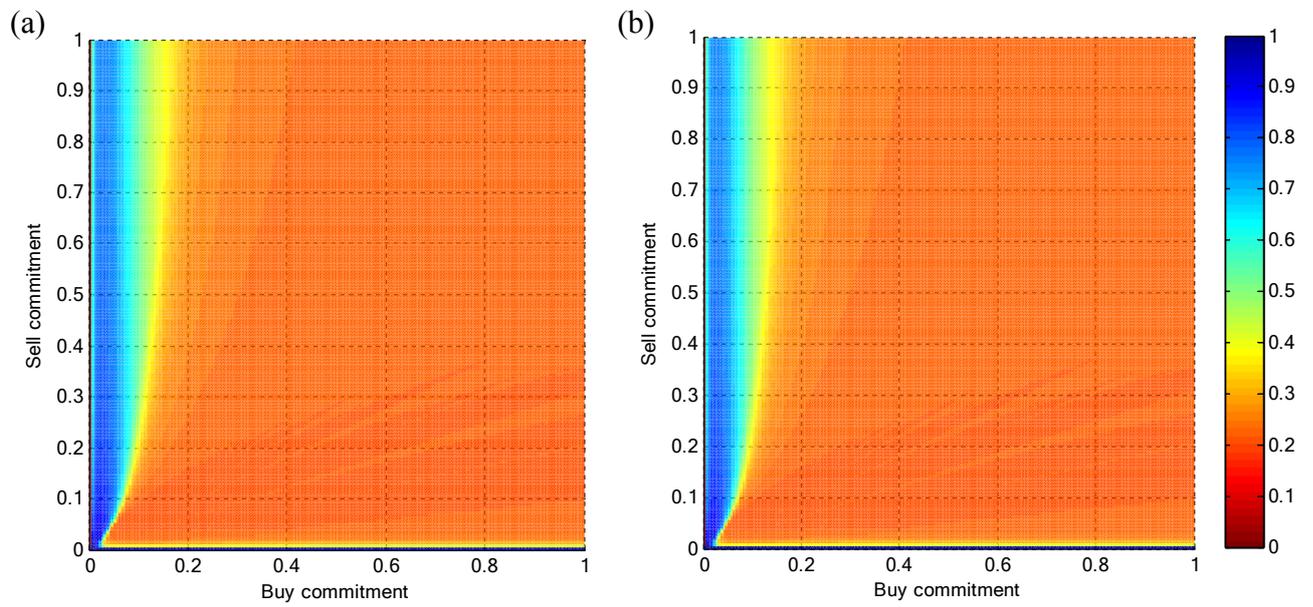

**Fig. S3: Observed Mo-thresholds for a price-crash at different buy and sell commitment parameters, comparing trading at (a) old prices and (b) new prices.** The difference is almost indiscernible. Chart (a) here is reproduced from Fig. 3b. As in Fig. 3, the buy commitments $k^+$ and sell commitments $k^-$ are each kept identical between Val and Mo.

**Appendix 3: Summary of simulations with multiple valuations**

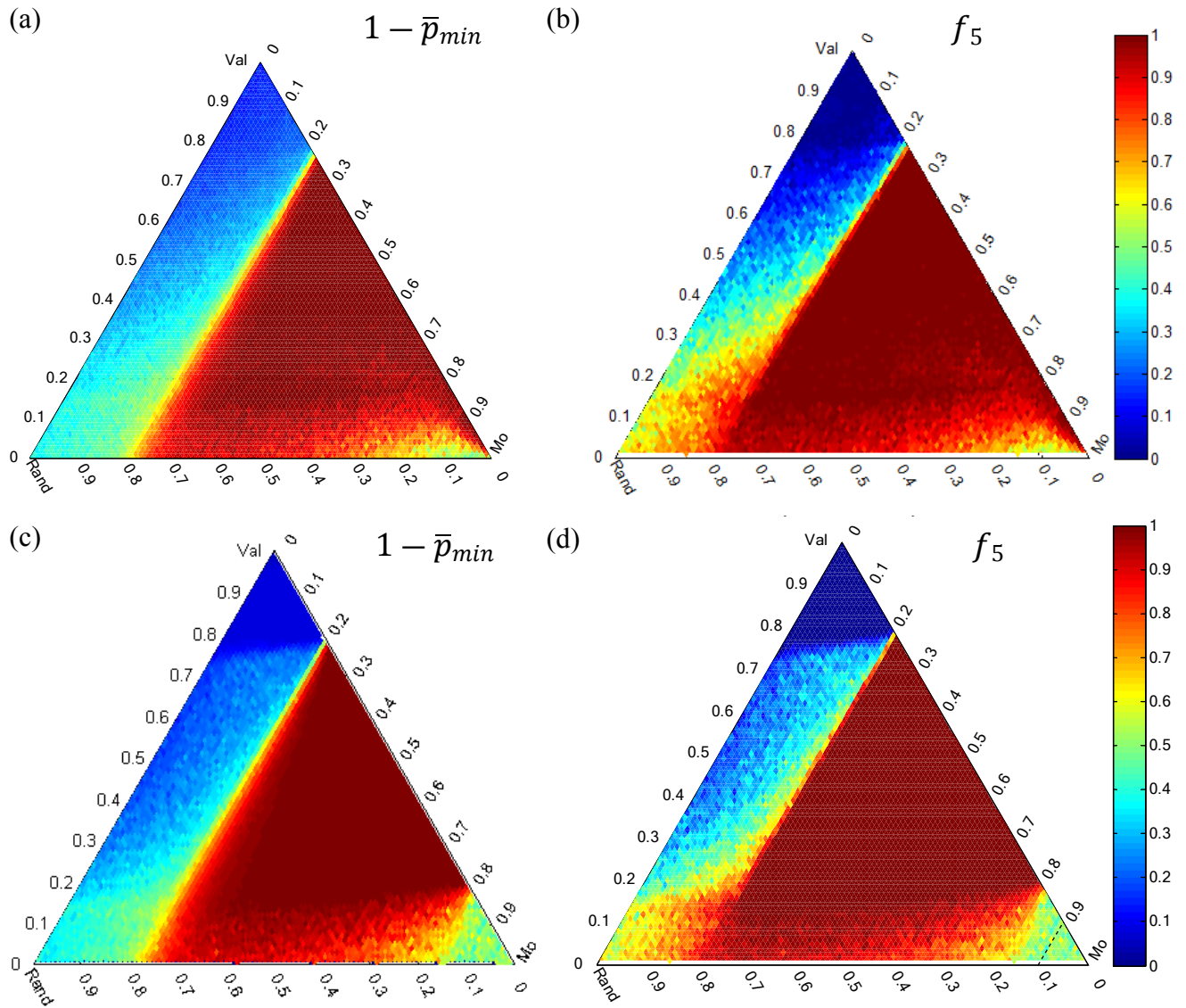

**Fig. S4: Price drops (relative to starting value) with different initial proportions of Val, Mo and Rand traders when there are 10 Vals: (a) relative price decrease; (b) frequency of a 30% decrease. For convenience, (c) and (d) reproduce Fig. 4 (single Val).** The top corner of the triangle represents 100% Val, the right corner 100% Mo and the left corner 100% Rand; a further 5047 starting points are calculated between these extremes, with 20 replicate simulations run from each starting point. Dark blue indicates no decline below 1, while dark red indicates (in a and c) large mean declines or (in b and d) predictable crashes, defined as a 30% drop in price over the 100 replicates. All simulations ran for 250 time-steps and used an asset–cash ratio of 4, momentum-smoothing constant of 0.002, buy and sell commitments of 10% and zero initial momentum.

**Appendix 4: Statistical properties of an estimator for τ**

Here we suggest an estimator for the tracking error based on a set of valuations. The estimator is shown to be consistent, converging almost surely to the true tracking error τ provided that the valuations are unbiased estimates of the true value $v$. It is also shown to be asymptotically normally distributed under weak conditions.

Suppose that there are $n$ valuation-based traders with independent unbiased estimates $\hat{u}_1,\ldots,\hat{u}_n$ of $u$. These can be thought of as independent random draws from an arbitrary distribution of mean $u$ and standard deviation $\sigma$. A better estimate than any single $\hat{u}_i$ is the unweighted mean of them all, denoted by $\hat{u}$. Provided that $\sigma < \infty$, $\hat{u}$ has variance that decays at rate $1/n$ as $n$ grows large. That is, in hypothetical replication, $\hat{u}$ is within $\frac{c}{\sqrt{n}}$ of $u$ with high probability, where $c$ is a constant.

Write $\log(x)$ for the natural logarithm of $x$ and:

$$\tau(x) = \frac{(\{\log(p/x)\}^2)^{1/2}}{\log 2}$$

so that $\underline{\tau} \triangleq \tau(u) = |\log_2(p) - \log_2(u)|$ and $\hat{\underline{\tau}} \triangleq \tau(\hat{u}) = |\log_2(p) - \log_2(\hat{u})|$ is an estimator of $\underline{\tau}$. Assume that $v$ and $\log(p/v)$ are bounded away from zero. By a Taylor series expansion of $\tau(x)$ around $u$, evaluated at $\hat{u}$,

$$\hat{\underline{\tau}} - \underline{\tau} = \sum_{k=1}^{\infty} \frac{(-1)^k (\hat{u}-u)^k \underline{\tau}}{k v^k \log(p/u)} = -\frac{(\hat{u}-u)\underline{\tau}}{v \log(p/u)} + O_{pr}(n^{-1}),$$

where $O_{pr}(n^{-1})$ is a term that tends in probability to zero at rate $n^{-1}$ as $n\to\infty$.

This implies by a central limit theorem for $(\hat{u}-v)\sqrt{n}$ that the leading term in an asymptotic approximation, with approximation error of order $1/\sqrt{n}$, to the density function of $\hat{\underline{\tau}}$ at $t$ is

$$f_{\hat{\underline{\tau}}}(t) \simeq \phi\left\{\frac{(t-\underline{\tau})\underline{\tau}\sigma}{v \log(p/u)\sqrt{n}}\right\},$$

where $\phi(x)$ is the standard normal density function at $x$.

The increase in orders of magnitude from errors of order $1/n$ to errors of order $1/\sqrt{n}$ is from the use of the central limit theorem. The distribution of the estimator can be specified to higher orders of accuracy under particular distributional assumptions on the valuations. For instance, suppose the $\hat{u}_i$ are gamma distributed of shape $\alpha > 0$ and rate $\beta > 0$ so that their density function is

$$f_{\hat{u}_i}(u) = \left\{\frac{\beta^\alpha}{\Gamma(\alpha)}\right\} u^{\alpha-1} e^{-\beta u}, u \geq 0,$$

and $v = \mathbb{E}(\hat{u}_i) = \alpha/\beta$. One may write

$$\hat{\underline{\tau}} - \underline{\tau} = \frac{\tau}{\log(p/u)} - Z + O_{pr}(n^{-1}),$$

where $Z = n^{-1} \Sigma_i k \hat{u}_i$ and $k = \tau/\{v \log(p/v)\}$. The moment generating function for $Z$ is $M_Z(t) = \{1 - tk/(n\beta)\}^{n\alpha}$, showing that $Z$ is gamma distributed of shape $n\alpha$ and rate $n\beta/k$. This implies, inter alia, that the finite sample bias of $\hat{\underline{\tau}}$ for $\underline{\tau}$ is at most of order $n^{-1}$, since $\mathbb{E}(Z) = \alpha k/\beta = \tau/\log(p/v)$.

The example with gamma-distributed valuation estimates is chosen because of its convenient properties under scaling and summation. For modelling valuations that must be strictly positive and are expected to cluster below the mean with some extreme values, it is at least a plausible distribution.